# An accurate nucleic acid-small molecule docking framework via geometric deep learning with large-scale pretraining

Shi Li[1,†], Xujun Zhang[1,†], Mingquan Liu[2,†], Hui Zhang[1,4], Shuoying Jia[1,4], Yu Kang[1,4], Tingjun Hou[1,3]*, Peichen Pan[1,3]*

[1]College of Pharmaceutical Sciences, Zhejiang University, Hangzhou 310058, Zhejiang, P. R. China

[2]Faculty of Health Sciences, University of Macau, Macau SAR, China

[3]Zhejiang Provincial Key Laboratory for Intelligent Drug Discovery and Development, Jinhua Institute of Zhejiang University, Jinhua 321299, Zhejiang, China

[4]Shanghai Innovation Institute, Shanghai 200231, China

[†]Equivalent authors

**Abstract**

Nucleic acids, including DNA and RNA, have emerged as critical therapeutic targets, complementing traditional protein-focused drug discovery. Modulating nucleic acid function through small-molecule interactions enables the regulation of gene expression and disease pathways. Molecular docking serves as a key computational technique for accelerating nucleic acid-directed drug discovery by predicting binding modes and prioritizing candidate compounds. However, conventional physics-based docking methods for nucleic acid targets suffer from limited accuracy and computational efficiency, while emerging deep learning approaches are fundamentally constrained by the scarcity of experimentally resolved nucleic acid-small molecule complex structures. Here we present NucleoDock, which, to our knowledge, is among the first deep learning frameworks specifically designed for nucleic acid-small molecule docking. NucleoDock addresses these challenges through three integrated strategies. First, we employ physics-guided large-scale pretraining on millions of docking-generated synthetic complexes, followed by fine-tuning on curated experimental co-crystal structures, to overcome data scarcity. Second, we introduce a dual-resolution, sequence- and structure-informed representation that fuses nucleotide-level descriptors—including primary sequence embeddings and secondary-structure topology—with atomistic three-dimensional features, enabling the model to capture both biological context and geometric detail. Third, we adopt a mixture density network (MDN)-based geometric scoring head that parameterizes a conditional probability distribution over interaction distances, supporting probabilistic pose evaluation and ranking. On an external benchmark of 125 nucleic acid-ligand complexes, NucleoDock achieves a top-1 pose prediction success rate of 56% at RMSD < 2.0 Å, far surpassing rDock (29%) while generating 100 candidate poses in approximately 5 seconds per complex. Virtual screening evaluation on the ROBIN benchmark also demonstrates better early enrichment. NucleoDock represents a step toward bridging the methodological gap between protein- and nucleic acid-directed computational drug discovery.

## Introduction

Nucleic acids have increasingly emerged as a critical class of therapeutic targets, complementing traditional protein-based drug discovery. Within the central dogma, DNA and RNA act upstream of proteins to orchestrate gene expression, making their modulation a direct means of reprogramming protein synthesis and influencing disease-related pathways. Notably, approximately 85% of the human genome is transcribed into RNA, yet only ~1.5% encodes polypeptides, underscoring the vast and largely untapped regulatory landscape of non-coding RNAs[1]. A diverse range of nucleic acid structures—including duplex DNA, G-quadruplexes, riboswitches, long non-coding RNAs, and expanded repeats—have been shown to be druggable by small molecules[2-6]. The clinical success of RNA-targeted therapeutics such as Risdiplam[7] and the established use of DNA-directed agents in oncology[5,8,9], further validate nucleic acids as mainstream drug targets[2,10]. To systematically explore this target space, high-throughput screening platforms originally developed for proteins have been adapted for nucleic acids[11,12], yet these approaches remain prohibitively expensive, labor-intensive, and difficult to scale, motivating the development of complementary computational strategies.

In recent years, nucleic acid-small molecule docking has remained dominated by classical structure-based approaches[13]. Existing methods are composed mainly of two lines: general-purpose docking engines originally developed for protein-ligand systems and subsequently applied or adapted to nucleic acid targets, such as AutoDock Vina[14], DOCK[15], Glide[16], GOLD[17], and rDock[18]; and a smaller group of nucleic-acid-oriented methods that modify sampling or scoring for RNA/DNA systems, including RLDOCK[19], and NLDock[20]. Nucleic acid-small molecule docking is fundamentally challenged by the intrinsic properties of nucleic acid targets, including pronounced conformational flexibility, highly charged backbones, dependence on electrostatics, metal ions, and hydration, and binding pockets that are often shallow, solvent exposed, and highly polar. These features make both sampling and scoring substantially more difficult than in protein-ligand docking, and recent benchmark studies confirm that pose-prediction accuracy remains only moderate across current methods[21]. At the same time, experimentally resolved nucleic acid-small molecule complex structures and benchmark datasets remain scarce[22-24], which severely limits the development and generalization of data-driven docking models. Consequently, although deep learning has become a major force in protein-ligand modeling, no deep-learning-dominant framework has yet emerged for nucleic acid-small molecule docking; in this field, deep learning studies remain concentrated mainly on virtual screening, binding-site and interaction prediction[25-29].

In this study, we present NucleoDock, the first deep learning framework capable of performing nucleic acid-small molecule docking and virtual screening. NucleoDock is built upon three tightly integrated design principles. First, at the data level, we address the scarcity of experimentally resolved nucleic acid-ligand complexes through physics-guided large-scale pretraining: the model is first trained on a substantially expanded corpus of docking-generated complexes and subsequently fine-tuned on curated experimental co-crystal structures, thereby recalibrating predictions toward experimentally realistic binding geometries. Second, at the representation level, NucleoDock employs a dual-resolution, sequence- and structure-informed encoding scheme that combines nucleotide-level descriptors—including primary sequence and secondary-structure features—with atomistic three-dimensional representations; nucleotide-level embeddings are projected onto atom-level features via a residue-to-atom mapping module, providing structure-aware biological context for pose scoring and refinement. Third, as an end-to-end system, NucleoDock seamlessly integrates pose generation, geometric refinement, and scoring within a unified pipeline, directly producing candidate binding conformations and ranking them for downstream applications. In benchmark evaluations, NucleoDock achieves substantial improvements in pose prediction accuracy, improving the top-1 success rate at RMSD < 2.0 Å from 29% (rDock) to 56%, corresponding to an absolute gain of 27 percentage points over established physics-based docking tools, while also demonstrating non-trivial virtual screening capability on nucleic acid targets.

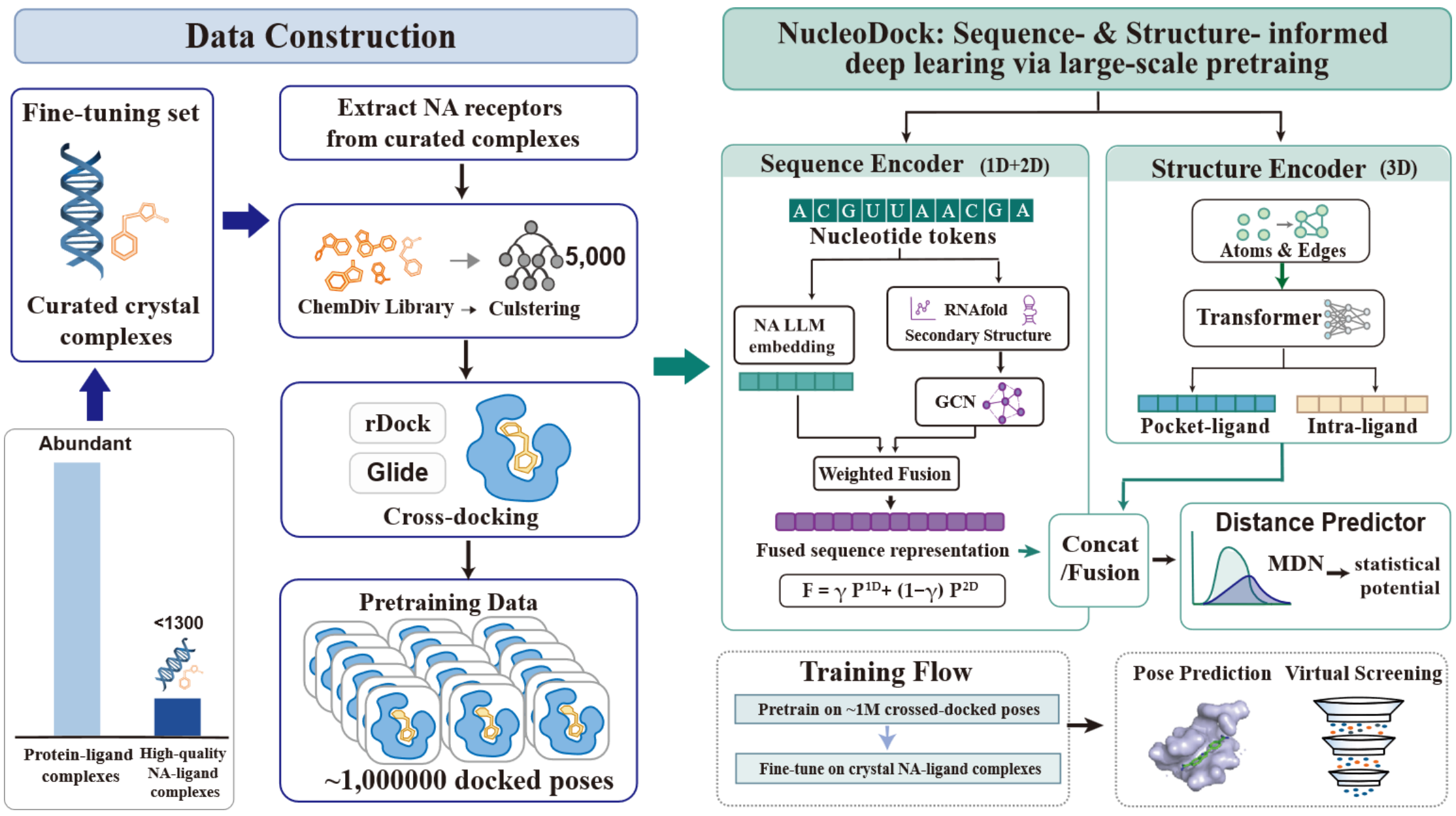

**Fig. 1 Overview of the NucleoDock framework**. NucleoDock combines large-scale cross-docking pretraining with crystal-complex fine-tuning to learn sequence- and structure-informed nucleic acid-ligand representations. The framework constructs approximately one million docked poses from curated nucleic acid receptors and diverse ligands, then integrates nucleotide sequence, secondary-structure, and 3D pocket-ligand features through dual encoders. The fused representation is used by an MDN-based distance predictor for pose prediction and virtual screening.

## Results

### Overview of NucleoDock

Fig. 2 illustrates the overall architecture of NucleoDock, which comprises two tightly coupled components: a multimodal feature extraction module and a geometry-aware prediction module. At the structural level, the ligand and nucleic-acid receptor are represented using single-atom features together with pairwise relational features derived from interatomic geometry. These representations are processed by separate encoders for the ligand and receptor, each taking atomic features and distance matrices as input. Within this framework, pairwise interactions are iteratively updated through a triangle-update attention mechanism, enabling the model to capture both intermolecular ligand-pocket interactions and intramolecular ligand geometry in a unified manner. The resulting ligand and receptor representations are subsequently integrated by an interaction encoder to form a joint structural cross-representation of the complex.

To complement atomic structural information with receptor-specific biological context, NucleoDock further incorporates an independent nucleic-acid encoder that processes the full-length RNA/DNA receptor through two parallel branches. A multilayer perceptron (MLP) encodes pretrained one-dimensional sequence embeddings derived from RNA-FM[30] or Nucleotide Transformer[31], whereas a graph convolutional network (GCN) captures two-dimensional topological dependencies. The outputs of these two branches are fused by learnable linear projections and weighted summation to generate context-aware residue-level representations. To align these residue features with the atomic interaction space, a residue-to-atom mapping module is introduced, allowing residue-level NA context to be projected back onto pocket atoms and concatenated with the structural pair representation.

The fused representation is then used for prediction directly at the atom-pair level. Specifically, the NA-enhanced ligand-pocket pair representation is transformed into intermolecular distance predictions, while a parallel branch predicts intra-ligand(holo) distances to preserve ligand geometry. In addition, the same fused ligand-pocket interaction representation is processed by an scoring head, which converts fine-grained pairwise interaction features into docking-quality scores and associated mixture-density parameters. In this way, distance prediction and pose scoring are jointly learned from a shared multimodal interaction space. To further reduce geometric distortion and enforce physically plausible conformations, the predicted coordinates are refined through learned translation, torsion, and rotation operations. The refined structures are then re-evaluated by the distance prediction module to generate ten refined distance matrices, which are combined with ten initial RDKit conformers to produce 100 candidate ligand poses for final ranking with GenoScore.

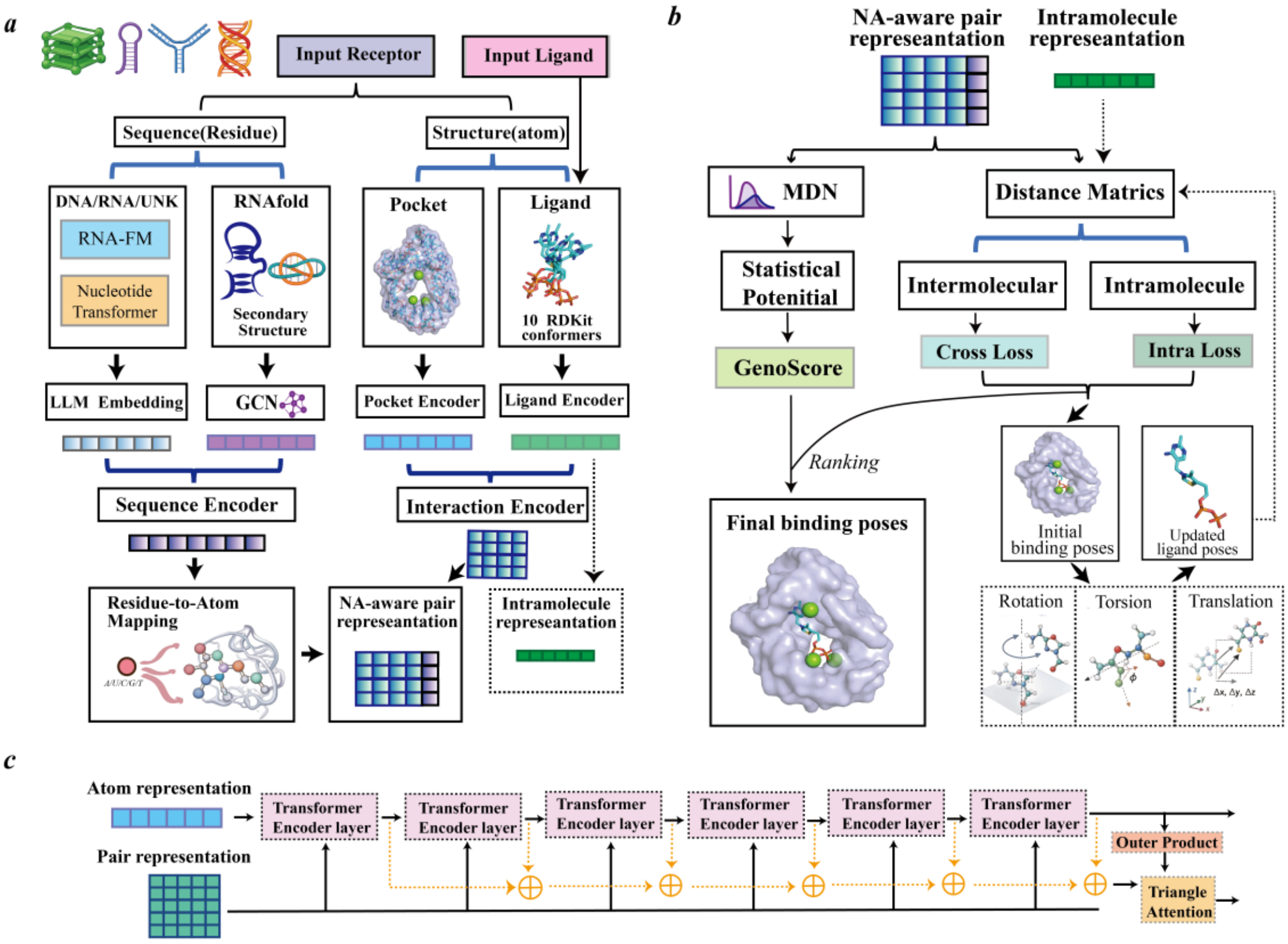

**Fig. 2 Model Architecture of NucleoDock. a) Multimodal feature extraction.** The receptor is encoded from residue-level sequence embeddings (RNA-FM or Nucleotide Transformer) and secondary-structure-derived topology (RNAfold followed by a GCN), producing context-aware residue representations. In parallel, ligand and pocket structures are encoded at the atomic level by a ligand encoder and a pocket encoder and integrated by an interaction encoder to yield structural cross and intra-ligand representations. Residue-level receptor features are aligned to pocket atoms via a residue-to-atom mapping module and injected into the structural cross representation to form an NA-aware pair representation. **b) Geometry prediction, refinement, and ranking.** Distance heads predict intermolecular (ligand-pocket) and intra-ligand (holo) distance matrices, supervised by cross-distance and holo-distance losses during training. In parallel, an MDN-based scoring head (i.e., GenoScore) parameterizes a statistical potential used for pose scoring. Predicted geometry is refined through learned translation, rotation, and torsion updates with optionl recycling, and candidate poses generated from refined predictions and initial predicted conformers are ranked by GenoScore to obtain the final binding poses. **c) Interaction Encoder.** A Transformer encoder with Evoformer-like pairwise refinement. Each layer takes a head-wise attention bias and produces attention maps that are aggregated into a pair representation; TriAtten then jointly updates token and pair features under padding masks, outputting refined tokens, updated pair features, a bias delta, and norm statistics for stability monitoring or auxiliary regularization.

## Re-Docking Performance

We evaluated NucleoDock on an external re-docking benchmark constructed by aggregating and deduplicating the four benchmark sets used by NLDock[18,32-34], yielding 125 non-redundant PDB complexes. Under a unified local semi-flexible docking protocol, NucleoDock achieved the best overall performance among all compared methods. Using the standard success criterion of top-1 RMSD < 2.0 Å, and counting failed docking attempts as failures, NucleoDock reached a top-1 success rate of 56% over all 125 complexes (Table 1). It also consistently outperformed the physics-based docking baselines across multiple RMSD cutoffs for the top-ranked pose (≤ 0.5, 1.0, 2.0, 2.5, and 5.0 Å; Fig. 3e). On each of the four original benchmark subsets (Yan, Ruiz, Chen, and Phiplis), NucleoDock further achieved the highest top-1 and top-5 success rates under the RMSD≤2.0 Å criterion (Fig. 3h,i; Table 2). Representative RNA and DNA re-docking examples are shown in Fig. 3a.

The baseline methods exhibited non-negligible failure rates on this benchmark: rDock, NLDock, and Glide failed to return a valid docking result for 6, 5, and 16 complexes, respectively. Because these failed cases were counted as unsuccessful predictions, the benchmark reflects overall practical docking robustness in addition to pose accuracy. Under this evaluation, NucleoDock remained the strongest method overall.

To assess whether performance was dominated by a small number of receptor families, we performed receptor structural comparison using RMalign/RMscore and clustered receptors at an RMscore threshold of 0.75. The 125 targets partition into 67 clusters, including 54 singletons (Fig. 3f), indicating substantial structural diversity. Although three larger clusters contain 19, 17, and 7 receptors, respectively, all remaining clusters are absent from the training set. After excluding these three dominant clusters, NucleoDock still achieved a top-1 success rate of 48.14% at RMSD < 2.0 Å, supporting its ability to generalize beyond the major structural families represented in training.

NucleoDock was also substantially more efficient. Averaged over the 125-complex benchmark, it generated 100 docked poses in approximately 5 s wall-clock time per receptor-ligand pair, markedly faster than rDock, NLDock, and Glide under the same timing protocol (Fig. 3j). These results indicate that NucleoDock improves both docking accuracy and computational efficiency.

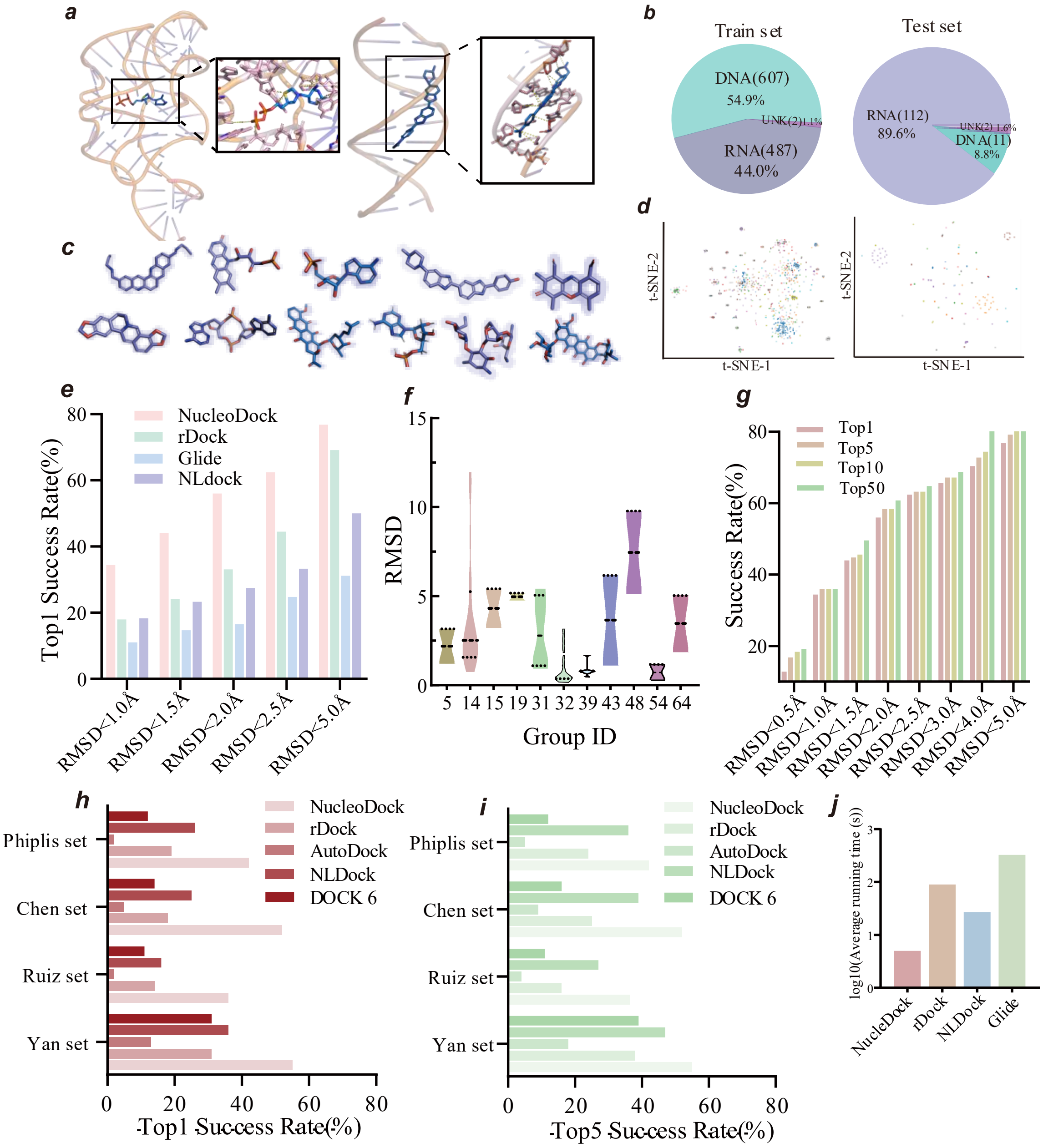

**Fig. 3 Dataset characterization and docking accuracy of NucleoDock.** (a) Representative NucleoDock-predicted complex structures for PDB IDs 2GDI and 1QV8, rendered in PyMOL. (b) Distribution of nucleic-acid types, including RNA, DNA, and unknown entries, in the training and external test sets. (c) Representative ligand examples from the training and validation sets. (d) Structural clustering of PDB entries from the training, validation, and external test sets using RMalign. (e) Top-1 NucleoDock docking success rates evaluated across multiple RMSD cutoffs: <0.5, <1.0, <2.0, <2.5, and <5.0 Å. (f) RMSD distribution across receptor structural clusters in the 125-complex external test set. (g) Top-k docking success rates of NucleoDock for k = 1, 5, 10, and 50, where success is defined as at least one pose within the top-k predictions achieving RMSD ≤1.0, ≤1.5, ≤2.0, ≤2.5, or ≤5.0 Å. (h, i) Comparison of top-1 and top-5 binding-pose prediction success rates, respectively, among NucleoDock, NLDock, AutoDock, rDock, and DOCK 6 on four benchmark subsets, using RMSD ≤2.0 Å as the success

criterion. (j) Average docking time per ligand across nucleic-acid complexes in the 125-complex external test set.

**Table 1.** Docking Performance Comparison: Success Rates of Top-1 Scored Conformations at Multiple RMSD Cutoffs (<1.0-5.0 Å) in 125 test set (NA complexes).

| | RMSD<1.0Å | RMSD<1.5 Å | RMSD<2.0 Å | RMSD<2.5 Å | RMSD<5.0 Å | avgRMSD |
|---|---|---|---|---|---|---|
| NucleoDock | **34.4%** | **44.0%** | **56.0%** | **62.4%** | **76.8%** | **2.95** |
| rDock | 15.1% | 22.7% | 29.4% | 38.7% | 66.4% | 3.89 |
| NLDock | 18.3% | 23.3% | 27.5% | 33.3% | 50.0% | 5.16 |
| Glide | 11.0% | 14.7% | 16.5% | 24.8% | 31.2% | 5.57 |

**Table 2.** Docking Performance Comparison: Success Rates of Top-1 and Top 5 Scored Conformations at Multiple RMSD Cutoffs (<1.0-5.0 Å) in 4 test set (Yan set, Ruiz set, Chen set, Philips set). The results from NLDock, AutoDock, rDock and DOCK 6 are obtained from Feng et al[20].

| **Toolkit** | **Top1 Pose** | | | | **Top5 Pose** | | | |
|---|---|---|---|---|---|---|---|---|
| | **Yan** | **Ruiz** | **Chen** | **Philips** | **Yan** | **Ruiz** | **Chen** | **Philips** |
| NucleoDock | **54.83%** | **37.84%** | **51.92%** | **41.67%** | **56.45%** | **41.50%** | **54.14%** | **44.44%** |
| NLDock | 36.36% | 16.07% | 25.00% | 26.19% | 46.75% | 26.79% | 39.29% | 35.71% |
| Autodock | 12.99% | 1.79% | 5.36% | 2.38% | 18.18% | 3.57% | 8.93% | 4.76% |
| rDock | 31.17% | 14.29% | 17.86% | 19.05% | 37.66% | 16.07% | 25.00% | 23.81% |
| Dock6 | 31.17% | 10.71% | 14.29% | 11.90% | 38.96% | 10.71% | 16.07% | 11.90% |

**Physical Plausibility of Docked Conformations**

To complement RMSD-based accuracy, we evaluated top-1 pose plausibility using PoseBusters, which reports pass/fail outcomes for chemical validity/consistency, intramolecular validity, and intermolecular validity[35]. This analysis is particularly relevant for AI-based docking methods, which can achieve favorable RMSD while still producing geometrically distorted or physically implausible ligand conformations[35]. In NucleoDock, such failure modes are mitigated by a geometric reconstruction stage that progressively refines ligand rotation, translation, and torsions with re-inference after each adjustment; representative re-docking examples across four PDB complexes (Fig. 4i) show that the reconstructed poses preserve structural integrity and avoid the nonphysical deformations often seen in unconstrained generative docking outputs.

We further analyzed robustness across sequence-similarity regimes by stratifying test structures according to their maximum sequence identity to the training set. Specifically, for each test structure, we aligned its sequence against all training structures using Parasail global alignment (Needleman--Wunsch) for binning. Although Buttenschoen et al.[35] used the partitions 0--30%, 30--95%, and 95--100%, the 0--30% subset in our evaluation was too small for reliable analysis; we therefore used the bins (0,70]%, (70,95]%, and (95,100]%. As shown in Fig. 4c, NucleoDock achieves relatively balanced performance across these identity ranges.

Overall, most top-1 predictions remain chemically consistent and intramolecularly plausible, with pass rates of 85.6% for chemical validity/consistency and 92.8% for intramolecular validity (Supplementary Table 7). Intermolecular validity is lower at 61.7% (Supplementary Table 8), indicating that residual receptor--ligand steric infeasibility remains the dominant failure mode. Because PoseBusters was developed primarily for protein--ligand complexes, these receptor-contact checks should be interpreted conservatively in the nucleic acid--ligand setting.

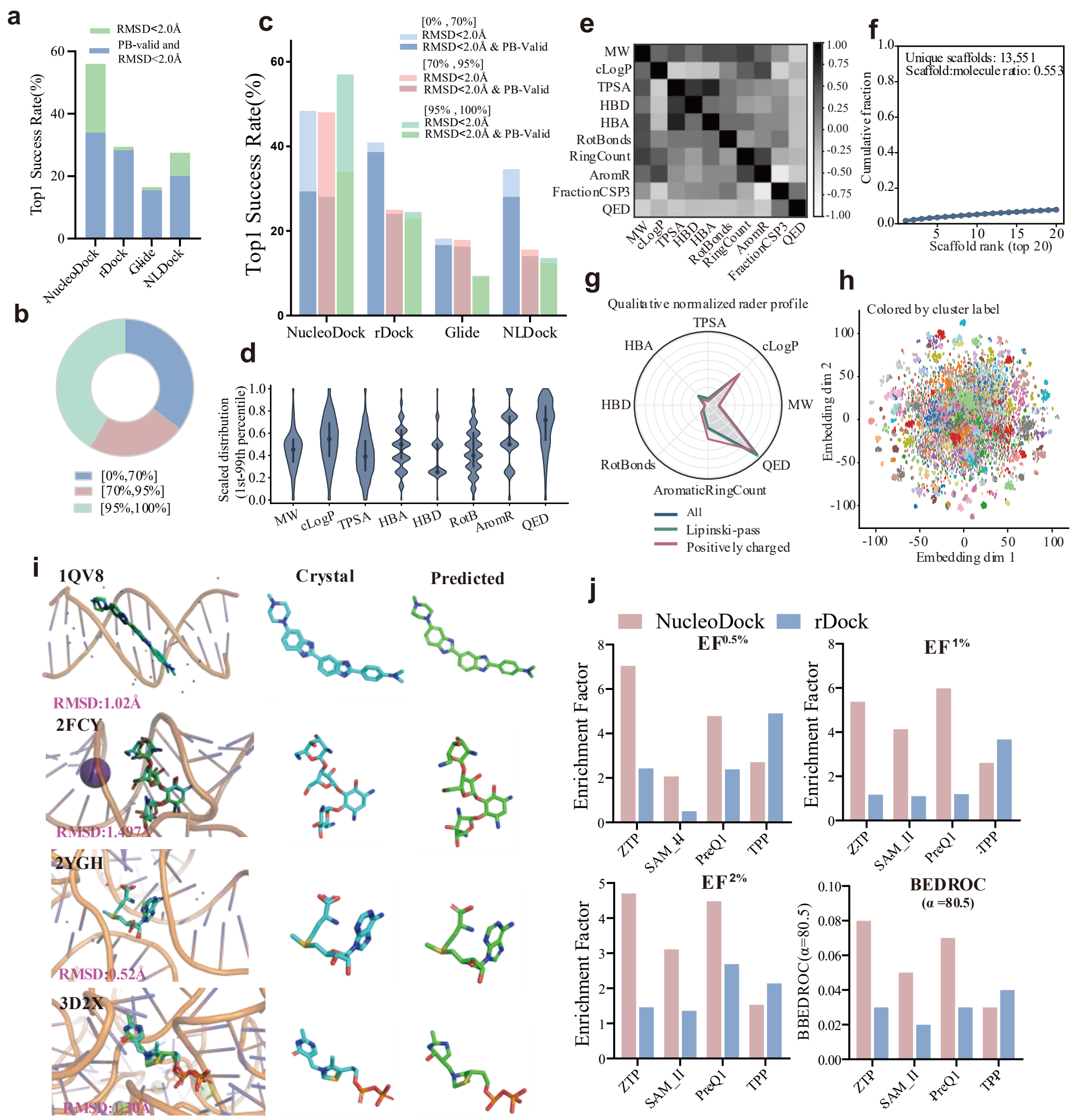

**Fig. 4. Pose-quality assessment, screening-library characterization, and virtual-screening performance.** (a-c) PoseBusters-based evaluation of top-ranked docking poses. (a) Overall top-1 PoseBusters success rates across docking programs. (b) Distribution of test complexes stratified by maximum sequence identity to the training set. (c) Top-1 PoseBusters success rates within three sequence-identity bins: (0,70]%, (70,95]%, and (95,100]%. (d-h) Physicochemical and chemical-space characterization of the ROBIN screening library. (d) Normalized distributions of key physicochemical descriptors, including molecular weight, cLogP, TPSA, HBA, HBD, rotatable bonds, aromatic rings, and QED, shown as violin plots with median and interquartile-range overlays. (e) Descriptor correlation heatmap. (f) Normalized radar plot comparing median descriptor profiles of the full library, Lipinski-compliant molecules, and a positively charged high-QED subset. (g) Scaffold diversity summarized by cumulative coverage of top-ranked Bemis-Murcko scaffolds. (i) Representative comparisons between NucleoDock-predicted poses and experimentally resolved structures. (j) Early enrichment and BEDROC-based virtual-screening performance of NucleoDock and rDock.

**Virtual Screening Performance**

We next evaluated NucleoDock in retrospective virtual screening on the external ROBIN benchmark using four structurally resolved RNA targets[26,27],: PreQ1 riboswitch (PDB 3FU2), SAM-II riboswitch (PDB 2QWY), ZTP riboswitch (PDB 5BTP), and TPP riboswitch (PDB 2GDI). We first characterized the ROBIN screening library to assess its suitability for nucleic-acid ligand screening. Among 24,572 input entries, 24,570 were RDKit-valid molecules, and 24,493 unique compounds remained after duplicate removal. The library was broadly distributed within drug-like small-molecule chemical space, with most compounds showing moderate molecular weight and cLogP values, while retaining a limited number of high-mass, highly polar, or highly lipophilic outliers. Fingerprint-based embedding and Bemis-Murcko scaffold analysis further revealed substantial chemical diversity, with 13,551 unique scaffolds and no dominant chemotype. Most compounds passed Lipinski and Veber filters, and aromatic or heteroatom-containing scaffolds were common, consistent with chemical features frequently involved in RNA ligand recognition.

The four selected targets cover diverse riboswitch binding-site architectures, including compact nucleobase recognition, SAM-dependent pseudoknot recognition, ZMP/ZTP-sensing two-domain recognition, and thiamine-pyrophosphate recognition. This benchmark therefore tests docking-based screening across RNA pockets with distinct topology, charge environment, and ligand chemistry. Using early-enrichment metrics, NucleoDock achieved EF values greater than 1 across all targets and cutoffs (Fig. 4j). It outperformed rDock on ZTP, SAM-II, and PreQ1 across all reported metrics, whereas rDock showed stronger early enrichment on TPP at EF0.5%, EF1%, and BEDROC. Although absolute BEDROC values were modest, as expected for the highly imbalanced ROBIN benchmark with a baseline BEDROC of 0.016, NucleoDock improved over rDock for most target-metric pairs. These results support the utility of NucleoDock for early-hit prioritization in nucleic-acid-targeted virtual screening.

**Discussion**

Recent advances in chemical biology have established nucleic acids, particularly structured RNAs and regulatory DNA elements, as increasingly tractable therapeutic targets. Yet, in contrast to the rapid progress of machine learning in protein-ligand docking and virtual screening, deep learning methods explicitly developed for nucleic acid-small molecule docking remain limited. This gap is consequential because nucleic acid recognition involves pronounced conformational heterogeneity, non-canonical geometries, and challenging electrostatic environments that complicate both sampling and scoring.

Here, we take a step toward this challenge by combining large-scale synthetic pretraining with fine-tuning on curated experimental co-crystal structures, and by introducing a sequence- and structure-informed framework for nucleic acid-ligand pose generation and scoring. Across external benchmarks, NucleoDock improves Top-1 pose prediction over established physics-based docking methods such as rDock, while also showing meaningful early enrichment in retrospective virtual screening on ROBIN. In particular, NucleoDock outperforms rDock on three of the four evaluated targets across all reported early-enrichment metrics, indicating that its learned scoring signals can improve the prioritization of true binders near the top of the ranked list. At the same time, the gains are not uniform across targets: TPP remains a challenging case in which rDock performs better at the strictest early-retrieval cutoffs, suggesting that target-specific binding characteristics are not yet captured equally well. Additionaly, although NucleoDock improves top-1 pose selection, its top-50 success rate remains below that of rDock, indicating insufficient near-native pose coverage during candidate generation.While the reconstruction stage reduces major geometric artifacts, some generated poses remain physically implausible, suggesting the need for tighter integration of sampling, steric constraints, and receptor-aware filtering.These limitations are closely linked to the current data regime of nucleic acid docking. Experimentally resolved nucleic acid-ligand structures remain scarce relative to protein-ligand complexes, limiting supervision for pose refinement, score calibration, and cross-target generalization. Although synthetic pretraining partially alleviates this constraint, it cannot fully capture the diversity and subtle physicochemical features of experimental complexes. Progress will therefore likely depend on both richer structural datasets and more effective integration of sparse experimental data with large computationally generated corpora.

From a practical perspective, NucleoDock is a structure-based method and thus requires a three-dimensional receptor model and a predefined search region. For targets lacking experimental structures, modeled receptor conformations may be used[36-38], although uncertainty in receptor geometry can propagate to docking predictions; in such cases, multiple receptor hypotheses may be beneficial. In addition, NucleoDock assumes a predefined search region and requires specification of a pocket center and/or docking volume, consistent with common grid-/cavity-based docking workflows. When binding-site annotations are unavailable, the search region can be initialized using homologous complexes, conserved functional motifs, or emerging RNA-small molecule binding-site prediction methods.

Overall, our results show that deep learning can meaningfully improve top-1 nucleic acid-ligand pose prediction, while also clarifying the main next steps: better near-native pose coverage, stronger score calibration for screening, and broader experimental structural support. Progress along these directions should strengthen the computational basis for rational small-molecule design against nucleic acid targets.

## Methods

### Data Set Preparations

The dataset construction in this study comprises two complementary components: (i) a crystal complex dataset used for supervised fine-tuning, and (ii) a large-scale docking dataset employed for model pretraining.

We first curated a fine-tuning dataset of nucleic acid-small molecule crystal complexes. Relevant structures were primarily collected from the NAKB[39] and PDB[40] databases and subsequently subjected to stringent filtering and manual curation. The selection criteria were as follows: (1) crystal structures with a resolution better than 2.5 Å; (2) ligands containing more than five atoms; and (3) in systems where both proteins and nucleic acids interact with the ligand, cases in which nucleic acids were manually determined to play the dominant role in ligand binding were retained. After applying these criteria, removing redundant entries, and excluding all complexes reserved for the external test sets, we obtained a total of 1,106 nucleic acid-small molecule complexes for model training. To prevent structural leakage between splits, we adopted a structure-based clustering protocol. We computed pairwise structural similarities for 1,106 RNA/DNA targets using RMalign[41] and formed clusters by thresholding at RMscore $\geqslant$ 0.75 and taking connected components. We then performed a cluster-level partition so that each cluster was assigned entirely to either training or validation, yielding a 7:1 train/validation split and ensuring that structurally similar targets were not shared across splits. To facilitate fair comparison with prior work, the 125 nucleic acid-small molecule complexes introduced in the NLDock[20] study were explicitly excluded from the training set and instead reserved as an independent test set in this work.

For pretraining-set construction, we assembled a drug-like ligand pool by combining nucleic-acid-associated compounds from ChEMBL[42] and ligands co-crystallized with RNA/DNA in the Protein Data Bank (PDB), yielding ~30,000 candidate molecules. To promote chemical diversity and reduce redundancy, we applied MaxMin diversity selection under a fingerprint-based Tanimoto distance criterion, selecting 5,000 representative compounds[43]. To mitigate scaffold leakage into evaluation, we then removed molecules belonging to the same chemical classes as those in the test set, leaving 4,962 ligands for pretraining.

To generate large-scale synthetic nucleic acid-ligand complexes for pretraining, we employed two physics-based docking engines, rDock or Glide. For rDock, receptor preparation and docking followed the standard rDock workflow, with the intermolecular scoring configuration specified via the dock_solv.prm setting; ligand preparation was performed in Maestro (details below). We removed docking outputs with anomalously high scores indicative of non-physical placements or failed runs. After filtering, the resulting rDock-derived corpus was partitioned by receptor overlap with the supervised dataset: 2,079,514 complexes whose receptors appeared in the supervised training set were used for pretraining, while 255,900 complexes were reserved for validation.

Motivated by recent benchmarking indicating that, with appropriately defined (tight) grid boxes, Glide SP can achieve nucleic acid-ligand docking performance comparable to or exceeding rDock in RNA-focused settings, we additionally generated a Glide-based docking corpus in standard precision mode. For each of the 1,106 binding sites, we defined an inner grid of $10 \times 10 \times 10$ Å and an outer grid of $25 \times 25 \times 25$ Å and performed systematic cross-docking between binding sites and the 4,962 ligands. After quality filtering, we retained only the top-scoring pose per prepared receptor-ligand state, resulting in 4,341,657 docked complexes for pretraining.

All receptor and ligand preparation steps were performed with the Schrödinger 2024 suite. Ligands were prepared using LigPrep, with ionization and tautomer states enumerated by Epik at pH 7.0 (±2.0), retaining up to four states per ligand. Receptors were processed using the Protein Preparation Wizard: protonation states were assigned/optimized at pH 7.0 using PROPKA, missing side chains were repaired, and structures were subjected to restrained energy minimization under the OPLS4 force field.

## Model Architecture

In summary, the framework comprises two distinct yet integrated subsystems: a Feature Extraction Module for representation learning, and a Distance Predictor Module for geometric inference.

$$P(\text{Binding}) = f(\text{Molecule,Pocket} \mid \text{Sequence})$$

This project introduces a multi-modal geometric deep learning framework for NA-ligand docking, specifically designed to enhance docking accuracy by integrating explicit NA topological and evolutionary semantics. The framework operates on a dual-stream fusion mechanism:

**3D Atomic Stream:** The backbone utilizes Transformer to encode the 3D coordinates and element types of the ligand and the rigid NA pocket atoms, capturing local geometric constraints.

**1D/2D Semantic Stream:** To incorporate global context, designed an independent NA encoder. This module processes the full-length NA sequence using two parallel branches: an MLP for 1D sequence embeddings (derived from RNA-FM or Nucleotide Transformer) and a GCN for 2D topological features. The outputs are fused via learnable linear projections and weighted summation to generate context-aware residue representations.

## Structure-based Encoder

The model implements a Transformer encoder augmented with an Evoformer-like pairwise refinement module[44] (Figure 2C). Given token embeddings of shape [B, L, C], it first applies layer normalization and embedding dropout and optionally zeroes out padded positions. The core backbone is a stack of encoder layers standard Transformer encoder blocks, each consuming an external attention bias (typically shaped as head-wise pair features) and returning attention probabilities. Across layers, the model accumulates these attention maps into a pair representation, effectively integrating multi-layer attention patterns with the initial bias. After replacing -inf entries with zeros and reshaping to [B, L, L, H], the model feeds both the token representation and the pair representation into a TriAtten module[45], which jointly updates tokens and pairwise features under token/pair masks derived from padding. The updated pair representation is then converted back into a head-major bias tensor and used to compute a delta-pair update relative to the input bias. Finally, the module outputs refined token features, the updated pair features, the pairwise delta, and two norm-based regularization statistics for tokens and pair deltas, enabling stability monitoring or auxiliary losses.

**Feature extraction**

For each complex, pocket residues were defined as those containing at least one atom within 6.0-8.0 Å of any ligand atom; pocket atoms were filtered to retain common elements (C, H, N, O, S, P) and annotated with a backbone/side-chain indicator. Pocket geometry was further summarized by a center and spatial extent computed from pocket coordinates $\mathbf{X}$ as $\mathbf{c} = (max(\mathbf{X}) + min(\mathbf{X}))/2$ and $\mathbf{s} = max(\mathbf{X}) - mean(\mathbf{X})$. Ligands were standardized using RDKit, and an ensemble of $M = 100$ candidate conformers was generated via distance-geometry embedding followed by MMFF optimization; conformers were aligned on heavy atoms and clustered using KMeans to select $N = 10$ representative conformations per ligand, with duplication-based padding when fewer valid conformers were available. To map atom-type annotations to contiguous integer indices, we predefined a vocabulary comprising 26 common ligand atom types and 6 nucleic-acid atom types, augmented with four special tokens: $[CLS]$ to denote sequence start, $[SEP]$ to denote sequence end, $[UNK]$ for out-of-vocabulary atom types, and $[PAD]$ for length normalization. For nucleic-acid sequence information, we obtained sequence representations directly from RNA-FM and the Nucleotide Transformer and appended a three-dimensional indicator encoding whether the sequence corresponds to DNA, RNA, or an unknown category (UNK).

**Encoder Architecture**

To effectively capture the structural intricacies of the complex, we employ distinct, independent encoders for the ligand and the pocket, respectively. $\mathbf{H}$ **denote the atom representation** and $\mathbf{Z}$ denote the pair representation.

$$\mathbf{H}_{\mathbf{m}}^{(\mathbf{0})}, \mathbf{Z}_{\mathbf{m}}^{(\mathbf{0})} = \text{MolEncoder}(E_{mol}) \quad (1)$$

$$\boldsymbol{H}_{\boldsymbol{p}}^{(\mathbf{0})}, \boldsymbol{Z}_{\boldsymbol{p}}^{(\mathbf{0})} = \text{PocketEncoder}(E_{pkt}) \quad (2)$$

The architecture of each encoder leverages a Transformer-based backbone to extract atom-level features. A critical innovation in this design is the construction of the initial pair representation through the accumulation of multi-head attention maps from the Transformer layers. Subsequently, to facilitate structural refinement, we introduce a Triangular Attention (TriAtten) mechanism. This module performs a co-evolutionary update of both atom and pair representations—analogous to the structural information update mechanism found in AlphaFold—yielding the final refined representations, $\mathbf{H}_{final}$ and $\mathbf{H}_{final}$

**Encoding Process**

The encoding process begins with feature initialization via normalization and dropout:

$$\boldsymbol{H_0} = \text{Dropout}(\text{LayerNorm}(\boldsymbol{E})) \tag{3}$$

Assuming a Transformer depth of $L$ layers, the sequential update for the $l$-th layer and the accumulation of pair information are defined as:

$$\boldsymbol{H}_l, \boldsymbol{A}_l = \text{TransformerLayer}(\boldsymbol{H}_{l-1}, \text{bias} = \boldsymbol{M_{init}}) \tag{4}$$

$$\boldsymbol{Z_{accum}} = \boldsymbol{M_{init}} + \sum_{l=1}^{L} \boldsymbol{A}_l \tag{5}$$

where $\mathbf{A}_l$represents the attention maps from the $l$-th layer. The final structural refinement is performed by the TriAtten module:

$$\boldsymbol{H}_{final}, \boldsymbol{Z}_{final} = \text{TriAtten}(H_L, Z_{accum}) \tag{6}$$

Following independent encoding, the representations are concatenated to form a unified feature space. We construct a block matrix $\mathrm{Z}_{concat}^{(0)}$where the diagonal blocks retain the self-pair representations from the individual encoders ($\mathrm{Z}_m$ for ligand, $\mathrm{Z}_p$for pocket). The off-diagonal blocks, representing the intermolecular interactions, are initialized using a mean-difference heuristic ($\phi$) to approximate spatial relationships:

$$\boldsymbol{Z_{concat}^{(0)}} = \begin{pmatrix} \boldsymbol{Z_m^{(0)}} & \phi\left(\boldsymbol{Z_m^{(0)}}, \boldsymbol{Z_p^{(0)}}\right) \\ \phi\left(\boldsymbol{Z_p^{(0)}}, \boldsymbol{Z_m^{(0)}}\right) & \boldsymbol{Z_p^{(0)}} \end{pmatrix} \tag{7}$$

The mean-difference function $\phi$ is defined as follows, capturing the relative feature deviations between the two modalities:

$$\phi(\boldsymbol{A}, \boldsymbol{B})_{ij} = \text{mean}(\boldsymbol{A_{i,:}}) - \text{mean}(\boldsymbol{B_{:,j}}) \tag{8}$$

To enable deep information exchange between the ligand and the pocket, the model utilizes a Concat Encoder to perform $K$ iterations of **recycling**. This iterative process allows the model to progressively refine the binding geometry.

$$\boldsymbol{H}^{(\boldsymbol{k})}, \boldsymbol{Z}^{(\boldsymbol{k})} = \text{ConcatEncoder}\left(\boldsymbol{H}^{(\boldsymbol{k-1})}, \boldsymbol{Z}^{(\boldsymbol{k-1})}\right), \quad k = 1 \ldots K \tag{9}$$

**Cross-Pair Representation Extraction:**

After recycling, we extract the cross-pair features (corresponding to ligand-pocket interactions) from the final decoded vector (denoted as **dec**). We enforce symmetry on these features to ensure physical consistency:

$$\boldsymbol{Z}_{\boldsymbol{cross}}[i, j] = \frac{1}{2}\left(\boldsymbol{Z}^{\boldsymbol{dec}}_{\boldsymbol{i},\boldsymbol{N_m}+\boldsymbol{j}} + \left(\boldsymbol{Z}^{\boldsymbol{dec}}_{\boldsymbol{N_m}+\boldsymbol{j},\boldsymbol{i}}\right)^T\right) \tag{10}$$

**Sequence-based Encoder**

**GCN Construction**

The Graph Convolutional Network is constructed by treating NA residues as nodes and their structural relationships as edges. Node Features: Initialized with pre-trained language model embeddings (771-dim) to capture evolutionary information. Edge Construction: Connectivity is derived from RNAfold predictions. Edges represent both the covalent backbone connections and non-covalent hydrogen bonds (base pairings) defined by the secondary structure. This allows the GCN to propagate information across the global fold, distinguishing stems, loops, and hairpins.

**Residue-to-Atom Mapping**

Unlike standard methods that use global pooling (which dilutes local signals), we implemented a precise Residue-to-Atom Mapping. By pre-calculating the alignment between PDB atoms and sequence indices, the model dynamically "gathers" the fused 1D/2D residue features and assigns them to the corresponding constituent atoms within the 3D pocket. This effectively "paints" the rigid 3D geometric input with rich, chemically relevant topological labels, enabling the docking head to recognize binding hotspots based on both geometry and secondary structure semantics.

**Multi Feature Fusion:**

In the final stage, we synthesize a comprehensive feature vector $\mathbf{F}_{fused}$ by integrating the interaction information with extrinsic biological context. This involves fusing the symmetric cross-pair representation, the decoded single-atom representations for both ligand and pocket, and the external **Nucleic Acid (NA) sequence embeddings**. This multi-modal fusion is mathematically expressed as:

$$\boldsymbol{F_{fused}}[i,j] = \left[ \underbrace{\boldsymbol{Z_{cross}}[i,j]}_{\text{Pair}} \parallel \underbrace{\boldsymbol{H_m^{dec}}[i]}_{\text{Mol}} \parallel \underbrace{\boldsymbol{H_p^{dec}}[j]}_{\text{Pocket}} \parallel \underbrace{\boldsymbol{E_{na}}}_{\text{NA}} \right] \tag{11}$$

Here, $\parallel$denotes the concatenation operation. This fused representation encapsulates pairwise geometric constraints, local atomic features, and global sequence information from the RNA/DNA, serving as the robust input for downstream scoring and distance prediction tasks.

**Distance Predictors:**

The fused interaction representation is converted into geometric predictions at the pair level rather than through global pooling. Specifically, for each ligand-pocket atom pair, we concatenate the refined cross-pair embedding, the decoded ligand atom embedding, and the decoded pocket atom embedding to form a cross-interaction tensor. NA-aware residue features are further mapped onto pocket atoms and appended to this tensor, yielding the final ligand-pocket representation used for prediction. An analogous holo representation is constructed for ligand-ligand atom pairs, augmented with a pooled NA context vector.

Two lightweight nonlinear heads are then applied to these fused representations. The first head predicts ligand-pocket distances, while the second predicts intra-ligand holo distances. In both branches, an ELU activation followed by a positive shift (+1) is used to ensure strictly positive distance outputs. The same fused ligand-pocket interaction tensor is also processed by a score head, which converts fine-grained pairwise interaction features into a scalar docking score.

**Loss function**

The loss function integrates geometric constraints from both intramolecular (ligand-ligand) and intermolecular (pocket-ligand) interactions. The distance loss is computed over a subset of valid atom pairs, defined by a mask $\mathcal{M}$, which filters for pairs with spatial distances below a predefined threshold.

$$\widehat{d_{ij}} = \mathrm{ELU}\left(\boldsymbol{W_{dist}} \cdot \boldsymbol{F_{fused}}[i,j] + \boldsymbol{b_{dist}}\right) + 1.0$$

$$\mathcal{L}_{dist} = \frac{1}{|\mathcal{M}|} \sum_{(i,j)\in\mathcal{M}} \mathrm{SmoothL1}\left(\widehat{d_{ij}}, d_{ij}^{GT}\right) \tag{12}$$

$$\text{SmoothL1}(\hat{d}, d) = \begin{cases} 0.5(\hat{d} - d)^2 & |\hat{d} - d| < 1 \\ |\hat{d} - d| - 0.5 & \text{otherwise} \end{cases}$$

For the confidence assessment, we employ a Mixture Density Network (MDN) loss, denoted as $\mathcal{L}_{MDN}$. For each atom pair, the model leverages the fused feature representation $\boldsymbol{F_{fused}}$ to regress the parameters of a Gaussian Mixture Model (GMM). Specifically, the network predicts the mixing coefficients ($\pi$), standard deviations ($\sigma$), and means ($\mu$) for the component distributions. The MDN loss is mathematically formulated as the Negative Log-Likelihood (NLL) of the ground truth distances under the predicted probability distribution. Furthermore, to address molecular symmetry during training, the loss function explicitly accounts for graph isomorphisms. This is achieved by evaluating all valid symmetric permutations of the ligand atoms and selecting the permutation that minimizes the calculated NLL.

$$\begin{aligned} \pi_k &= \text{Softmax}\left(\boldsymbol{W}_\pi \boldsymbol{F_{fused}} + \boldsymbol{b}_\pi\right)_k \\ \sigma_k &= \text{ELU}\left(\boldsymbol{W}_\sigma \boldsymbol{F_{fused}} + \boldsymbol{b}_\sigma\right) + 1.3 \\ \mu_k &= \text{ELU}\left(\boldsymbol{W}_\mu \boldsymbol{F_{fused}} + \boldsymbol{b}_\mu\right) + 1.0 \end{aligned} \tag{13}$$

$$\mathcal{L}_{MDN} = -\frac{1}{|\mathcal{M}|} \sum_{(i,j)\in\mathcal{M}} \log\left(\sum_{k=1}^{K} \pi_k \mathcal{N}\left(d_{ij}^{GT} \mid \mu_k, \sigma_k\right)\right) \tag{14}$$

$$\mathcal{L}_{Total} = \lambda_1 \cdot \mathcal{L}_{cross} + \lambda_2 \cdot \mathcal{L}_{intra} + \lambda_3 \cdot \mathcal{L}_{MDN} \tag{15}$$

**Scoring**

GenoScore is employed for pose selection during conformational prediction which could also be used for screening. The **MDN score** is defined as the summation of probability densities scaled by the inverse square of the distance ($1/d^2$). This metric quantifies interaction strength by mimicking the physical decay properties intrinsic to Coulombic and Van der Waals forces.

$$\text{MDNScore} = \sum_{(i,j)\in\mathcal{M}} \sum_{k=1}^{K} \frac{\pi_{ijk} \cdot \mathcal{N}\left(d_{ij} \mid \mu_{ijk}, \sigma_{ijk}\right)}{d_{ij}^2 + \epsilon} \tag{16}$$

$$\text{GenoScore} = \lambda \times (\mathcal{L}_{cross} + \mathcal{L}_{intra}) - \text{MDNScore}$$

$i$: Index of the ligand atom ($1 \ldots N_{lig}$)
$j$: Index of the pocket atom ($1 \ldots N_{pkt}$)
$k$: Index of the Gaussian mixture component ($1 \ldots K$).

**Geometry optimization**

To convert the network-predicted geometric constraints into physically plausible 3D poses while preserving ligand topology, we reconstruct ligand coordinates by optimizing only rigid-body motion and rotatable-bond torsions, rather than freely updating all atomic coordinates. Following the parameterization of Corso et al., each ligand conformation is represented by a (6+$m$)-dimensional vector

$$\mathrm{n} = [\mathrm{x}, \mathrm{y}, \mathrm{z}, \phi, \theta, \psi, \alpha_1, \alpha_2 \ldots, \alpha_m] \tag{17}$$

where $(x, y, z)$ denotes translation, $(\phi, \theta, \psi)$ are Euler angles defining rigid-body rotation, and $\{\alpha_k\}_{k=1}^{m}$ are dihedral angles of the $m$ rotatable bonds. Applying this low-dimensional transform via a deterministic kinematic operator ensures that bond lengths, bond angles, and ring geometries are preserved throughout optimization, thereby enforcing fundamental chemical constraints.

For each complex, we define a local coordinate system centered at the binding pocket. Specifically, we compute the pocket centroid $c_p$ as the mean of pocket atom coordinates and translate pocket coordinates by subtracting $c_p$. Each initial ligand conformer is likewise centered by subtracting its own centroid, placing ligand and pocket in a shared, origin-centered frame that initializes the ligand near the pocket center in translation space. After reconstruction, the optimized ligand coordinates are transformed back to the original global frame by adding $c_p$.

Given the predicted ligand-pocket distance constraints and intraligand distance constraints, we iteratively refine $\mathbf{n}$ by minimizing a distance-matching objective:

$$\mathcal{L}_{\mathrm{dist2coords}} = \mathcal{L}_{\mathrm{cross}} + \mathcal{L}_{\mathrm{lig}} = \mathrm{SmoothL1}(\Delta D_{cross}) + \mathrm{SmoothL1}(\Delta D_{lig}) \tag{18}$$

where $\Delta D_{\mathrm{cross}}$ and $\Delta D_{\mathrm{lig}}$ denote the discrepancies between the updated distances and the network-predicted distances for ligand-pocket and intraligand atom pairs, respectively. We optimize $\mathbf{n}$ using L-BFGS (step size $lr = 0.1$) with gradients computed by automatic differentiation. Optimization runs for at most $I$ outer iterations and employs early stopping with patience $P$: if the objective does not improve for more than $P$ consecutive iterations, optimization terminates and the best parameters (lowest loss) are retained. To avoid retaining poorly constrained reconstructions, we discard candidate solutions whose best loss exceeds a fixed threshold (best loss $\geq 100$).

To improve robustness to initialization and prediction noise, we use a multi-start strategy with $N_c$ RDKit-generated ligand conformers and test-time ensembling with $N_t$ augmented network predictions. For each pair of (initial conformer, constraint prediction), we run the above dist2coords optimization to obtain one reconstructed pose; in our default setting $N_c = 10$and $N_t = 10$, yielding $N_c \times N_t = 100$candidate poses per complex. Each reconstructed pose is further evaluated by an MDN-based geometric consistency score, and candidates are ranked by the composite criterion

$$\text{GenoScore} = \lambda \mathcal{L}_{\text{dist2coords}} - \text{MDNScore} \tag{19}$$

where $\mathcal{L}_{\text{dist2coords}}$(a SmoothL1 distance-matching objective) and MDNScore (a likelihood-based geometric consistency measure) have different numerical scales. The coefficient $\lambda$is therefore introduced to **balance the relative contribution** of the reconstruction loss and the probabilistic consistency term during pose ranking. We selected $\lambda = 5$using a coarse grid search on the validation set and fixed it across all experiments. Performance was **insensitive within a reasonable range** of $\lambda$, indicating that the method does not rely on fine-tuning this hyperparameter. With smaller GenoScore indicating higher confidence.

Because dist2coords constitutes the primary runtime bottleneck in high-throughput docking, we provide multiple execution backends, including (i) a CPU L-BFGS implementation, (ii) multi-process parallelization across candidates, and (iii) a GPU-accelerated backend (including a cuLBFGS-B-style solver) to reduce wall-clock time while preserving docking accuracy.

**Model Training**

We pre-trained the model on millions of docked complexes generated by rDock or Glide and subsequently fine-tuned it on crystallographic structures. The model architecture remains largely consistent between the pre-training and fine-tuning stages, with differences confined to training hyperparameters. Notably, sequence-level representations were not incorporated during the pre-training phase. The key model parameters are summarized as follows:

Table 3. The crucial model hyperparameter settings for NucleoDock

| Hyperparameters | Settings | | |
|---|---|---|---|
| | Pre-training | Fine-tuning | inference |
| Dimension of structure hidden representation | 768 | 768 | 768 |
| Dimension of Sequence hidden representation | \ | 128 | 128 |
| Number of layers for ligand encoder | 6 | 6 | 6 |
| Number of layers for pocket encoder | 6 | 6 | 6 |
| Number of attention heads(H) | 16 | 16 | 16 |
| Number of layers for interactive encoder | 6 | 6 | 6 |
| Number of recycles for interactive encoder | 3 | 3 | 3 |
| Threshold for the calculation of NA-ligand distance | 8 | 7 | 7 |
| Learning Rate | 1e-4 | 1e-4 | \ |
| Batch size | 64 | 16 | \ |
| Epoch | 5 | 80 | \ |
| Weight decay | 1e-4 | 1e-6 | \ |
| Ratio of warmup phase to the total phase | 0.05 | 0.05 | \ |
| Optimizer | AdamW | AdamW | LBGFS |
| Initial learning rate for warmup phase | 1e-7 | 1e-8 | \ |
| Weight of distance loss for NA-ligand atom pairs ($W_{cross}$) | 1 | 1 | 1 |
| Weight of distance loss for intramolecular pairs in ligand ($W_{intra}$) | 1 | 1 | 1 |
| Weight of MDN loss ($W_{MDN}$) | 1 | 1 | 1 |

## Ablation Study

We conducted an ablation study to quantify the contribution of several model components to docking accuracy on a benchmark set of 125 nucleic acid complexes. Ablation experiments systematically remove or disable individual model elements to assess their impact on performance, isolating the effect of each component while holding other factors constant, consistent with standard practice in deep learning evaluation.

In our experimental setup, we evaluated the following configurations: (i) Without Pre-training, where the model is trained only on experimentally resolved structures; (ii) Without Sequence (1D), omitting one-dimensional sequence features; (iii) Without Sequence (2D), omitting two-dimensional structural topology features; (iv) Without Sequence (1D + 2D), removing all sequence-derived representations; and (v) the full NucleoDock model. By comparing docking accuracy metrics across these variants, we quantify the individual and combined contributions of pre-training and sequence-informed features to overall performance.

Table 4. Impacts of several strategies on docking accuracy based on 125 test NA-complexes

| Strategy | Top1 success rates(%)(＜2Å) | avgRMSD |
|---|---|---|
| Without Pre-training | 46.4 | 3.38 |
| Without Sequence(1D) | 50.3 | 3.05 |
| Without Sequence(2D) | 52.1 | 3.14 |
| Without Sequence(1D+2D) | 47.3 | 3.15 |
| NucleoDock | 56.2 | 2.95 |

## External re-docking benchmark

We constructed an external evaluation benchmark by aggregating the four benchmark sets used in NLDock, namely the Yan set (77 complexes), Ruiz set (56 complexes), Chen set (56 complexes), and Phiplis set (42 complexes). Overlapping entries appearing in multiple sets were identified by PDB ID and removed, yielding a final non-redundant benchmark of 125 receptor-ligand complexes. All re-docking experiments for NucleoDock and baseline methods were performed on this unified benchmark

**Docking Protocols for Baseline Methods**

We compared NucleoDock against rDock, NLDock, and Glide under standardized baseline protocols. For rDock, we explicitly accounted for the fact that it does not perform intramolecular geometry relaxation during docking; specifically, bond lengths, bond angles, and ring conformations remain fixed throughout the search. Consequently, docking performance is highly sensitive to the quality and diversity of the input ligand conformers. To provide a conformer ensemble that was sufficiently diverse while remaining computationally tractable, we generated 100 RDKit conformers for each ligand and clustered them to select 10 representative conformations as docking inputs. Docking was then performed using the standard rDock workflow with a designated reference ligand, and the dock_solv.prm scoring configuration was enabled to better account for solvation and metal-ion effects that are often relevant to nucleic acid-ligand recognition. In preliminary experiments, using only a single RDKit conformer per ligand substantially degraded performance, reducing the rDock Top-1 success rate to approximately 5% under the RMSD < 2.0 Å criterion. For the virtual-screening experiments, rDock was run with its default settings using the dock.prm configuration.

NLDock was executed using the default docking settings provided by the original authors. For Glide, all docking experiments were conducted using Schrödinger 2024. Receptor and ligand structures were prepared in Maestro, and ligands were further processed with Epik at a target pH of 7.0 ± 2.0. Docking was subsequently performed in Glide SP mode with default settings. During grid generation, the inner and outer box sizes were set to (10, 10, 10) and (25, 25, 25), respectively. We observed that appropriately chosen inner and outer box sizes consistently improved docking performance, in agreement with prior report

**Runtime Measurement**

Runtime was measured under a unified timing protocol that started after receptor and ligand preprocessing and excluded ligand conformer generation. Baseline docking programs were run in their standard CPU-only docking mode with one thread per job, whereas NucleoDock used GPU-accelerated inference. For each receptor-ligand pair, we recorded the wall-clock time required to generate 100 docked poses. All timing experiments were performed on the same workstation equipped with Intel Xeon Platinum 8570 CPUs and an NVIDIA GeForce RTX 4090 GPU.

**Structural Clustering Analysis**

To assess whether the external benchmark was dominated by a small number of receptor families, we performed receptor-level structural comparison using RMalign/RMscore. Receptors were clustered using an RMscore threshold of 0.75, following prior RNA structural or binding-site clustering practice. We then examined the resulting cluster-size distribution and determined whether each cluster was represented in the training set. To evaluate generalization beyond dominant structural families, we additionally recomputed re-docking performance after excluding the three largest receptor clusters.

**Virtual-screening benchmark.**

We used ROBIN as an external benchmark for retrospective nucleic-acid virtual screening. Following prior ROBIN-based evaluations, we selected four representative targets: PreQ1 (PDB 3FU2), SAM_II (PDB 2QWY), ZTP (PDB 5BTP), and TPP (PDB 2GDI). To avoid information leakage, these PDB structures were excluded from the training and validation sets. Because practical virtual screening mainly depends on prioritizing top-ranked candidates, performance was assessed using early-enrichment metrics, including EF at 0.5%, 1%, and 2%, together with BEDROC.

**Virtual Screening Evaluation Metrics**

Consistent with prior studies, we assessed VS perfor mance primarily based on two well-recognized metrics, including Boltzmann enhanced discrimination of receiver operating characteristic (BEDROC, a = 80.5), and enrichment factors (EFs). BEDROC incorporates a weighting parameter a into AUROC to emphasize early recognition capability. Here, we set a = 80.5, meaning that the top 2%of ranked molecules contributed to 80% of the BEDROC score. $EF^{x\%}$,denoted as the ratio of true positives found in the top x% of ranked compounds relative to random selection, offers a more intuitive measure of early recognition. We evaluated EF at 0.5%,1%,and 2% thresholds to capture performance across different screening depths. When the ***EF*** value exceeds 1, the method demonstrates a significant enrichment of active compounds.

$$EF = \frac{\frac{Hit^{sampled}}{N^{sampled}}}{\frac{Hit^{total}}{N^{total}}}$$

$\boldsymbol{N^{sampled}}$ refers to the top N compounds ranked by score, and $\boldsymbol{Hit^{sampled}}$ represents the number of active compounds among the top N sampled compounds. $\boldsymbol{Hit^{total}}$ refers to the total number of active compounds in the database, and $\boldsymbol{N^{total}}$ denotes the total number of compounds in the entire database.

**Data and code availability**

All nucleic-acid complex structures and corresponding sequences used in this study are publicly available from the RCSB Protein Data Bank (PDB)( https://www.rcsb.org/). The ROBIN benchmark for virtual screening, including tables summarizing small-molecule microarray (SMM) screening results, is publicly available via its GitHub repository(https://github.com/ky66/ROBIN). Structural clustering was performed using RMalign (http://www.rnabinding.com/RMalign/RMalign.html). For sequence-based representations, we used RNA-FM (via GitHub at https://github.com/ml4bio/RNA-FM) and the Nucleotide Transformer v2 (250M) ((https://huggingface.co/InstaDeepAI/nucleotide-transformer-v2-250m-multi-species) to compute 1D sequence embeddings. Secondary-structure predictions were obtained using RNAfold (https://github.com/ViennaRNA/ViennaRNA/releases ). Source code, scripts, and execution details for our pipeline can be found at https://github.com/ShiYue3384/NucleoDock.

**Acknowledgments**

This work was financially supported by the National Key R&D Program of China (2025YFF0521500), the National Natural Science Foundation of China (82473843, 22503081), the Postdoctoral Fellowship Program of CPSF (GZC20252381), the China Postdoctoral Science Foundation (2024M762886), and the Central Guidance for Local Science and Technology Development Funds Project (2025ZY01022). We also thank the Information Technology Center and State Key Lab of CAD&CG, Zhejiang University for the support of computing resources.

## Supplementary Materials

Table.S1 The list of PDB structures from NAKB and PDB used for training are shown in the list below.

| 100D | 101D | 102D | 108D | 109D | 110D | 121D | 127D | 128D | 129D | 130D | 131D | 144D | 151D | 152D |
|---|---|---|---|---|---|---|---|---|---|---|---|---|---|---|
| 159D | 166D | 173D | 182D | 185D | 193D | 195D | 198D | 1AGL | 1AL9 | 1AMD | 1AO1 | 1AW4 | 1C9Z | 1CX3 |
| 1D11 | 1D14 | 1D17 | 1D21 | 1D22 | 1D30 | 1D32 | 1D36 | 1D37 | 1D38 | 1D43 | 1D44 | 1D45 | 1D46 | 1D48 |
| 1D58 | 1D63 | 1D64 | 1D67 | 1D85 | 1D86 | 1DA0 | 1DA9 | 1DCR | 1DJ6 | 1DL8 | 1DNE | 1DNH | 1DNS | 1DPL |
| 1DSD | 1DVL | 1EDR | 1EEL | 1EI4 | 1EM0 | 1ET4 | 1EVV | 1F6C | 1FD5 | 1FDG | 1FMQ | 1FMS | 1FN2 | 1FQ2 |
| 1FUF | 1G3X | 1G4Q | 1G5L | 1GJ2 | 1I0T | 1I1P | 1I2X | 1I2Y | 1I3W | 1I5V | 1IMR | 1IMS | 1JO2 | 1JTL |
| 1K2L | 1K2Z | 1K9G | 1KCI | 1KGK | 1KVH | 1L0R | 1L1H | 1L1V | 1LEJ | 1LEX | 1LEY | 1M69 | 1MP7 | 1MTG |
| 1N37 | 1N7A | 1N7B | 1NAB | 1NJN | 1OE5 | 1OFX | 1OLD | 1OVF | 1P9X | 1PIK | 1PQQ | 1PRP | 1PWF | 1Q9X |
| 1QSX | 1R68 | 1RAW | 1S2R | 1TN2 | 1UNJ | 1UNM | 1VRO | 1VS2 | 1VTG | 1VTH | 1VTI | 1VTJ | 1VZK | 1WOE |
| 1XCS | 1XVR | 1Y0Q | 1Y84 | 1Y8L | 1YLS | 1YYO | 1Z3F | 1Z58 | 1Z5T | 1Z8V | 1ZJG | 1ZPH | 1ZPI | 202D |
| 209D | 210D | 211D | 215D | 216D | 217D | 224D | 227D | 231D | 234D | 236D | 245D | 248D | 258D | 261D |
| 264D | 267D | 268D | 269D | 276D | 278D | 282D | 288D | 289D | 292D | 293D | 296D | 298D | 2A04 | 2A5R |
| 2ARG | 2B0K | 2B2B | 2B3E | 2CKY | 2D47 | 2DA8 | 2DBE | 2DCG | 2DND | 2DYW | 2EES | 2EET | 2EEU | 2EEV |
| 2ELG | 2ET0 | 2EZF | 2EZG | 2F8W | 2G5K | 2GB9 | 2GCV | 2GJB | 2GVR | 2GWA | 2GYX | 2H0W | 2H0Z | 2HO6 |
| 2HOJ | 2HOM | 2HOO | 2HOP | 2HRI | 2I2I | 2I5A | 2IE1 | 2KD4 | 2KMJ | 2KXM | 2L1V | 2L7V | 2L8H | 2LLJ |
| 2LWK | 2M4Q | 2MCC | 2MCO | 2MG8 | 2MGN | 2MIY | 2MS6 | 2MXS | 2N0J | 2N6C | 2NEO | 2NLM | 2NZ4 | 2O1I |
| 2O43 | 2O45 | 2OEY | 2OYQ | 2PIK | 2PL8 | 2R2U | 2RF3 | 2RSK | 2RU7 | 2TRA | 2YIE | 302D | 303D | 304D |
| 306D | 308D | 311D | 319D | 320D | 321D | 323D | 324D | 326D | 328D | 336D | 355D | 358D | 360D | 367D |
| 385D | 386D | 3AJK | 3AJL | 3B4A | 3B4B | 3B4C | 3BNQ | 3BNR | 3C3Z | 3C5D | 3C7R | 3CCO | 3CDM | 3CE5 |
| 3D2G | 3D2V | 3DIG | 3DIM | 3DIO | 3DIQ | 3DIR | 3DIX | 3DIY | 3DIZ | 3DJ0 | 3DJ2 | 3E5E | 3E5F | 3EGZ |
| 3EQW | 3ERU | 3ES0 | 3ET8 | 3EUI | 3EUM | 3EY0 | 3EY2 | 3EY3 | 3F2T | 3F2W | 3F2X | 3F2Y | 3F30 | 3F4E |
| 3F4H | 3FT6 | 3FWO | 3FX8 | 3G8T | 3G96 | 3G9C | 3GCA | 3GJH | 3GJJ | 3GSJ | 3GSK | 3GX6 | 3I1D | 3I5L |
| 3IQR | 3IRW | 3JQ4 | 3K1V | 3L3C | 3MIJ | 3MUT | 3MUV | 3N4O | 3NP6 | 3NX5 | 3NYP | 3NZ7 | 3OIE | 3OMJ |
| 3OZ5 | 3QCR | 3QF8 | 3QRN | 3QSC | 3QSF | 3RKF | 3SC8 | 3SKI | 3SKL | 3SKR | 3SKT | 3SKW | 3SKZ | 3SLQ |
| 3T5E | 3TD1 | 3TZR | 3U05 | 3U08 | 3U0U | 3U38 | 3U4M | 3UVF | 3UXW | 3UYB | 3UYH | 3V7E | 3WRU | 403D |
| 428D | 432D | 442D | 443D | 447D | 448D | 449D | 452D | 453D | 454D | 459D | 465D | 473D | 474D | 482D |
| 4AH0 | 4AH1 | 4B5R | 4BZT | 4BZU | 4BZV | 4D9X | 4D9Y | 4DA3 | 4DAQ | 4E1U | 4E7Y | 4E8K | 4E8M | 4E8N |
| 4E8Q | 4E8R | 4E8S | 4E8T | 4E8V | 4E8X | 4E95 | 4ERL | 4F8U | 4F8V | 4FAQ | 4FAR | 4FAU | 4FAW | 4FAX |
| 4FEJ | 4FEL | 4FEN | 4FEO | 4FEP | 4FRG | 4FRN | 4FXM | 4G0F | 4GMA | 4GPW | 4GPX | 4GPY | 4GQJ | 4GXY |
| 4HQI | 4I9V | 4IJ0 | 4ITD | 4JD8 | 4JIY | 4K31 | 4K32 | 4KZD | 4L5K | 4L81 | 4LTF | 4LTG | 4LTH | 4LTI |
| 4LTK | 4LTL | 4LVW | 4LVX | 4LVY | 4LVZ | 4LW0 | 4LX5 | 4LX6 | 4M3I | 4M3V | 4MJ9 | 4MS5 | 4NFO | 4NYA |
| 4NYC | 4NYD | 4NYG | 4O5W | 4O5X | 4OQU | 4P1D | 4P20 | 4P3S | 4P5J | 4P95 | 4PDQ | 4Q9Q | 4Q9R | 4QIO |
| 4QK9 | 4QKA | 4QLM | 4QLN | 4QVI | 4R4A | 4R8J | 4RE7 | 4RZD | 4TS0 | 4TS2 | 4TZX | 4TZY | 4U8A | 4U8B |
| 4W90 | 4W92 | 4X18 | 4X1A | 4XNR | 4XW7 | 4XWF | 4Y1I | 4Y1J | 4Y1M | 4YAZ | 4YB0 | 4YMC | 4ZC7 | 5BJO |
| 5C45 | 5C7U | 5C7W | 5CCW | 5CDB | 5D5L | 5DAM | 5DE5 | 5DE8 | 5DHB | 5ET2 | 5FK1 | 5FK4 | 5FK5 | 5FK6 |
| 5FKF | 5FKG | 5FKH | 5HBW | 5HIX | 5IP8 | 5IU5 | 5IWJ | 5JEU | 5JEV | 5JLW | 5KPY | 5KRG | 5KVJ | 5KX9 |
| 5LFS | 5LFW | 5LFX | 5LIG | 5LIT | 5LWJ | 5MVB | 5NPM | 5O62 | 5O69 | 5OB3 | 5ODF | 5ODM | 5OE1 | 5SWE |
| 5T83 | 5TPY | 5UED | 5UEE | 5UX3 | 5UZA | 5V0H | 5V0O | 5V1L | 5V3F | 5V9Z | 5VCF | 5VCI | 5VJ9 | 5VJB |
| 5XI1 | 5XJZ | 5XZ1 | 5Z1H | 5Z1I | 5Z71 | 5Z8F | 5ZEI | 5ZEJ | 6AST | 6AZ4 | 6BFB | 6BNA | 6BST | 6C63 |
| 6C65 | 6C8E | 6C8K | 6C8L | 6C8M | 6C8O | 6CB3 | 6CC1 | 6CC3 | 6CCW | 6CHR | 6CK4 | 6CK5 | 6D8A | 6DB8 |
| 6DLR | 6DLS | 6DLT | 6DMC | 6DMD | 6DME | 6DN1 | 6DN2 | 6DN3 | 6DY5 | 6E1S | 6E1U | 6E1V | 6E1W | 6E84 |
| 6E8T | 6E8U | 6FC9 | 6FZ0 | 6G7Z | 6G8S | 6GIM | 6GYV | 6GZ7 | 6GZK | 6GZR | 6HAG | 6HBT | 6HBX | 6HC5 |
| 6HWG | 6I4N | 6IJV | 6IJW | 6J0H | 6J2W | 6JBF | 6JBG | 6JJ0 | 6JJH | 6JJI | 6JWD | 6JWE | 6K3Y | 6KFI |
| 6KFL | 6KN4 | 6KXZ | 6LNZ | 6M4T | 6M5B | 6M5J | 6N5K | 6N5N | 6N5O | 6N5P | 6N5Q | 6N5S | 6N5T | 6O2L |
| 6P2H | 6P45 | 6PNK | 6PQ7 | 6Q57 | 6QIV | 6QN3 | 6R6D | 6RIO | 6RNL | 6RSO | 6RTI | 6S15 | 6S7D | 6SX3 |
| 6T3N | 6T3R | 6T3S | 6TB7 | 6TF0 | 6TF1 | 6TF2 | 6TF3 | 6TFE | 6TFF | 6TFG | 6TFH | 6U6J | 6U7Y | 6U7Z |
| 6U8F | 6U8U | 6UBU | 6UC7 | 6UC8 | 6UC9 | 6UP0 | 6V0L | 6V9D | 6VA2 | 6VA3 | 6VA4 | 6VMY | 6VUI | 6VWT |
| 6WTL | 6WTR | 6WZR | 6WZS | 6XB7 | 6XCL | 6XKN | 6XKO | 6XRQ | 6XT7 | 6YL5 | 6YLB | 6YMI | 6YMJ | 6YMK |
| 6YMM | 7A3Y | 7ATG | 7CPS | 7CSK | 7D12 | 7D5E | 7D7W | 7D7X | 7D7Y | 7D7Z | 7D81 | 7D82 | 7DFY | 7DJU |
| 7DJW | 7DWH | 7E9E | 7E9I | 7EAF | 7EDL | 7EDM | 7EDT | 7EL7 | 7ELP | 7ELQ | 7ELR | 7ELS | 7EOG | 7EOH |
| 7EOJ | 7EOK | 7EOL | 7EOM | 7EON | 7EOO | 7EOP | 7FGT | 7FHI | 7FJ0 | 7FVS | 7FVT | 7JNH | 7KBW | 7KBX |
| 7KU4 | 7KUK | 7KUL | 7KUM | 7KUN | 7KUO | 7KUP | 7KVT | 7KVU | 7KVV | 7KWK | 7L0Z | 7LNE | 7LNF | 7LNG |
| 7N7D | 7N7E | 7OA3 | 7OAV | 7OAW | 7OAX | 7OTB | 7PNG | 7Q7X | 7Q7Y | 7Q7Z | 7Q80 | 7Q81 | 7Q82 | 7QSH |
| 7RWR | 7SZU | 7TD7 | 7TDA | 7TDB | 7TDC | 7TUV | 7TZR | 7TZS | 7TZT | 7TZU | 7U87 | 7U88 | 7U89 | 7U8A |
| 7V9E | 7W9N | 7WG | 7X2Z | 7X3A | 7XLV | 7YEB | 7YEC | 7YEE | 7YEK | 7YEL | 7Z9P | 8ASH | 8B6K | 8BAR |
| 8C41 | 8C57 | 8C58 | 8C63 | 8C8J | 8CA7 | 8CAI | 8CF1 | 8CF8 | 8CGI | 8CMM | 8CXS | 8CXU | 8CXV | 8CXX |
| 8CXZ | 8CY1 | 8CY4 | 8CY5 | 8D33 | 8E85 | 8E86 | 8E87 | 8E88 | 8E8B | 8E8C | 8E8D | 8E8E | 8E8G | 8E8J |
| 8EC1 | 8ED6 | 8EDA | 8EDB | 8EYU | 8EYW | 8F1S | 8F1V | 8FB4 | 8FDQ | 8FDR | 8FHV | 8FHX | 8FNG | 8FNH |
| 8G2N | 8GXB | 8GXC | 8HB1 | 8HB8 | 8HZD | 8HZE | 8HZK | 8HZM | 8I3Z | 8ILD | 8ILF | 8ILH | 8JOZ | 8K7W |
| 8KEB | 8KED | 8KHH | 8OET | 8OJ7 | 8OY6 | 8OY7 | 8OY8 | 8OY9 | 8OYA | 8OYB | 8OYR | 8Q60 | 8QJK | 8QMB |
| 8QMH | 8RBJ | 8S84 | 8SCG | 8SCK | 8SCL | 8SCM | 8SCO | 8SCP | 8SCQ | 8SCR | 8SCS | 8SCT | 8SCU | 8SKW |
| 8SWO | 8SX5 | 8SX6 | 8SXL | 8SY1 | 8T8T | 8TAA | 8TC6 | 8U0P | 8U5J | 8U5T | 8U5Z | 8UJT | 8UJV | 8UJX |
| 8V7A | 8V7B | 8V7C | 8V7D | 8V7E | 8V7G | 8V7H | 8V7I | 8V7J | 8V7K | 8VAW | 8VAX | 8VB7 | 8VBG | 8VBH |

| 8VDH | 8VDI | 8VDN | 8VFA | 8VFD | 8VFH | 8VFI | 8VIU | 8VQV | 8VVJ | 8VZL | 8W7W | 8WQ7 | 8X21 | 8X22 |
|---|---|---|---|---|---|---|---|---|---|---|---|---|---|---|
| 8XP8 | 8XPB | 8XWU | 8XXG | 8XXK | 8XZE | 8XZL | 8XZM | 8XZN | 8XZO | 8XZP | 8XZQ | 8XZR | 8XZW | 8YAM |
| 8YNO | 9AVR | 9BDW | 9BS2 | 9BS4 | 9BUN | 9C3S | 9C3T | 9C51 | 9C9M | 9CHW | 9CI9 | 9DDR | 9DRC | 9DTS |
| 9E5J | 9E5K | 9E5L | 9E5P | 9E5Q | 9E5R | 9E5T | 9ELR | 9EMI | 9EWB | 9EWD | 9EWG | 9EWZ | 9F37 | 9FCO |
| 9FIB | 9FMF | 9FT8 | 9FZX | 9GCH | 9GGC | 9H8G | 9HDO | 9HDP | 9I22 | 9I9W | 9IWF | 9IWG | 9J1N | 9J1O |
| 9J6P | 9JL7 | 9K7R | 9K8Z | 9KKZ | 9KL1 | 9LJV | 9LKV | 9MOZ | 9MUJ | 9N26 | 9N39 | 9NKC | 9NKD | 9OL2 |
| 9SFQ | 9UW0 | | | | | | | | | | | | | |

Table.S2. The list of PDB structures from NAKB and PDB used for tests are shown in the list below.

| 1AJU | 1AKX | 1AM0 | 1ARJ | 1BYJ | 1DB6 | 1EHT | 1EI2 | 1F1T | 1F27 | 1FMN | 1FYP | 1I9V | 1J7T | 1KOC | 1KOD |
|---|---|---|---|---|---|---|---|---|---|---|---|---|---|---|---|
| 1LC4 | 1LVJ | 1MWL | 1NBK | 1NEM | 1NTA | 1NTB | 1NZM | 1O15 | 1O9M | 1P96 | 1PBR | 1Q8N | 1QD3 | 1QV4 | 1QV8 |
| 1R4E | 1TOB | 1UTS | 1UUD | 1UUI | 1XPF | 1Y26 | 1Y27 | 1YKV | 1YRJ | 1ZZ5 | 2AU4 | 2B57 | 2BE0 | 2BEE | 2D55 |
| 2ESI | 2ESJ | 2ET3 | 2ET4 | 2ET8 | 2F4S | 2F4T | 2F4U | 2FCX | 2FCY | 2FCZ | 2FD0 | 2G9C | 2GDI | 2GIS | 2JUK |
| 2JWQ | 2KGP | 2KTZ | 2KU0 | 2KX8 | 2L94 | 2MB3 | 2O3V | 2O3W | 2O3X | 2OE5 | 2OE8 | 2PWT | 2TOB | 2XNW | 2XNZ |
| 2XO1 | 2YDH | 2YGH | 2Z74 | 2Z75 | 316D | 3C44 | 3D2X | 3DIL | 3DS7 | 3DVV | 3E5C | 3F2Q | 3FO4 | 3FO6 | 3FU2 |
| 3G4M | 3GAO | 3GER | 3GES | 3GOG | 3GOT | 3GX2 | 3GX3 | 3GX5 | 3GX7 | 3LA5 | 3MUM | 3MUR | 3MXH | 3NPN | 3NPQ |
| 3Q3Z | 3Q50 | 3S4P | 3SD3 | 3SLM | 3SUX | 407D | 4AOB | 4ERJ | 4FE5 | 4JF2 | 4KQY | 4LVV | | | |

Table.S3. Comparison of average docking time per ligand across nucleic acid complexes in test set.

| | NucleoDock | rDock | NLDock | Glide |
|---|---|---|---|---|
| Average CPU time(s) | 5 | 90 | 27 | 327 |

*On a single workstation equipped with Intel Xeon Platinum 8570 CPUs (56 cores per socket, 2 threads per core; 112 physical cores/224 logical CPUs; 4 NUMA nodes) and an NVIDIA GeForce RTX 4090

Table.S4. Atomic Token Vocabulary Used by the Model

| Atom type of ligand | Atom type of receptor |
|---|---|
| C, N, O, S, H, Cl, F, Br, I, Si, P, B, Na, K, Al, Ca, Sn, As, Hg, Fe, Zn, Cr, Se, Gd, Au, Li | C, H, O, P, S, N |

Table S5**.** Early-stage enrichment and BEDROC of NucleoDock and rDock.

| Toolkit | | NucleoDock | | | | rDock | | | |
|---|---|---|---|---|---|---|---|---|---|
| Target | $n_{hits}$ | $EF^{0.5\%}$ | $EF^{1\%}$ | $EF^{2\%}$ | BEDROC ($\alpha$ =80.5) | $EF^{0.5\%}$ | $EF^{1\%}$ | $EF^{2\%}$ | BEDROC ($\alpha$ =80.5) |
| ZTP | 173 | 7.04 | 5.38 | 4.70 | 0.08 | 2.43 | 1.17 | 1.46 | 0.03 |
| SAM_ll | 194 | 2.07 | 4.14 | 3.11 | 0.05 | 0.5 | 1.10 | 1.36 | 0.02 |
| PreQ1 | 164 | 4.78 | 5.98 | 4.48 | 0.07 | 2.39 | 1.21 | 2.69 | 0.03 |
| TPP | 163 | 3.22 | 2.61 | 1.53 | 0.03 | 4.90 | 3.67 | 2.14 | 0.04 |

*Due to the low hit rate in the ROBIN dataset, the baseline BEDROC ( α =80.5) is 0.016 for each of the four target

Table.S6. Early-stage enrichment of AutoDock.

| Toolkit | | **AutoDock** | | | |
|---|---|---|---|---|---|
| Target | $n_{hits}$ | $\mathbf{EF^{0.5\%}}$ | $\mathbf{EF^{1\%}}$ | $\mathbf{EF^{2\%}}$ | **BEDROC ($\alpha$ =80.5)** |
| ZTP(5BTP) | 173 | 3.50 | 5.38 | 4.70 | 0.08 |
| SAM_ll(2QWY) | 194 | 0.00 | 0.00 | 0.00 | 0.05 |
| PreQ1(3FU2) | 164 | 3.58 | 2.39 | 2.69 | 0.07 |
| TPP2GDI | 163 | 2.45 | 3.67 | 3.37 | 0.03 |

Table S7. Physical Plausibility Comparison: PoseBusters pass rates for the top-1-scored conformations in the nucleic acid (NA) test set, reported for chemical validity/consistency,

intramolecular validity, and intermolecular validity.

| | Chemical validity & consistency | intramolecular validity | intermolecular validity |
|---|---|---|---|
| NucleoDock | 85.6 | 92.8 | 61.7 |
| rDock | 90.3 | 93.1 | 97.5 |
| Glide | 95.8 | 100.0 | 97.9 |
| NLDock | 80.6 | 96.4 | 70.8 |

*PoseBusters criteria that are evidently incompatible with the energy and chemical characteristics of nucleic acid-ligand systems were excluded from this evaluation. Without this adjustment, all docking methods would yield zero success rates on the physical plausibility metrics. For instance, certain ligands in nucleic acid complexes intrinsically contain radical species, a feature that is also observed in experimentally resolved complexes deposited in the PDB.

Table S8. Physical Plausibility Comparison: PoseBusters pass rates for the top-1-scored conformations in the nucleic acid (NA) test set, reported for Minimum protein-ligand distance, Minimum distance to organic cofactors, Volume overlap with organic cofactors and Volume overlap with protein.

| | Minimum protein ligand distance | Minimum distance to organic cofactors | Volume overlap with protein | Volume overlap with organic cofactors |
|---|---|---|---|---|
| NucleoDock | 61.7 | 100 | 86.3 | 100.0 |
| rDock | 98.3 | 98.3 | 97.5 | 99.17 |
| Glide | 100.0 | 100.0 | 100.0 | 97.98 |
| NLDock | 100.0 | 70.9 | 100.0 | 100.0 |